\documentclass[reprint,aps,groupedaddress]{revtex4-1}

\usepackage{graphicx}
\usepackage{amsmath}
\usepackage{amsfonts}
\usepackage{bbm}
\usepackage{physics}
\usepackage{xcolor}

\long\def\comment#1{}

\begin{document}
\title{Periodically driven three-dimensional Kitaev
  model}
  
 \author{Soumya Sasidharan}
 \author{Naveen Surendran}
\email[]{naveen.surendran@iist.ac.in}

\affiliation{Indian Institute of Space Science and Technology,
  Valiamala, Thiruvananthapuram-695547, Kerala, India}
  
\date{\today}

\begin{abstract}

We study the dynamics of a three-dimensional generalization of Kitaev's honeycomb lattice spin model (defined on the hyperhoneycomb lattice) subjected to a harmonic driving of $J_z$, one of the three types of spin-couplings in the Hamiltonian. Using numerical solutions supported by analytical calculations based on a rotating wave approximation, we find that the system responds nonmonotonically to variations in the frequency $\omega$ (while keeping the driving amplitude $J$ fixed) and undergoes dynamical freezing, where at specific values of $\omega$, it gets almost completely locked in the initial state throughout the evolution. However, this freezing occurs only when a constant bias is present in the driving, i.e., when $J_z = J'+ J\cos \omega t$, with $J'\neq 0$. Consequently, the bias  acts as a switch that triggers the freezing phenomenon. Dynamical freezing has been previously observed in other integrable systems, such as the one-dimensional transverse-field Ising model.

\end{abstract}


\maketitle
\include{reference.bib}

\section{\label{s-intro}Introduction}

Kitaev's honeycomb lattice model consists of spin-1/2 degrees of freedom located at the vertices of the lattice \cite{Kit(06)}. Nearest neighbor spins interact via anisotropic Ising-like couplings, with the anisotropy direction dependent on the orientation of the link between the spins. Remarkably, the model is exactly solvable and has a quantum spin liquid ground state. Quantum spin liquids (QSL) are characterized by the absence of any magnetic ordering while having long-range entanglement, which can give rise to elementary excitations with fractional quantum numbers \cite{ SavBal(16),TakTak(19)}. Depending on the relative strength of the spin interactions, Kitaev model exists in two distinct phases: a gapped phase supporting abelian anyons and a gapless phase supporting non-abelian anyons as elementary excitations. 

 Over the last few decades, experimental search for QSLs has identified several candidate materials. Examples include $\kappa$-$(\mathrm{BEDT}$-$\mathrm{TTF})_2\mathrm{Cu}_2(\mathrm{CN})_3$ \cite{ShiMiy(03)}, $\mathrm{ZnCu}_3(\mathrm{OH})_6\mathrm{Cl}_2$ \cite{FuIma(15)}, $\mathrm{NaYbSe}_2$ \cite{ZhuPan(23),DaiZha(21),JiaGon(20)}, and $\mathrm{NaYbS}_2$ \cite{WuLi(22)}. Since the discovery of Kitaev model, the search for QSLs has expanded to potential physical realizations of the model. Even though the strong and non-uniform anisotropy of the spin interactions in the model initially appeared unrealistic, Jackeli and Khaliullin showed that such interactions can arise from strong spin-orbit coupling and, in particular, become dominant in certain iridates \cite{JacKha(09), WinTsi(17)}. Following their proposal, new materials such as  $\mathrm{Na}_2\mathrm{IrO}_3$ \cite{SinGeg(10)}, $\alpha$-$\mathrm{Li}_2\mathrm{Ir}O_3$ \cite{SinMan(12)}, and $\mathrm{H}_3\mathrm{LiIr}_2\mathrm{O}_6$ \cite{KitTak(18)} have been synthesized and studied as possible realizations of two-dimensional Kitaev QSL. $\alpha$-RuCl$_3$ is another material that has been extensively studied as a strong candidate to be a Kitaev QSL \cite{BanBri(16),BanYan(17),KasOhn(18)}.  For comprehensive reviews of Kitaev materials, see Refs. \onlinecite{HerKim(18), TakTak(19), WenYu(19), TreHic(22)}.

Kitaev's construction has been extended to three dimensions based on various approaches \cite{Ryu(09), WuAro(09), ManSur(09)}. One particular extension involves defining the Hamiltonian on a hyperhoneycomb lattice and using the same type of interactions as in the original model \cite{ManSur(09)}. The hyperhoneycomb lattice model is also exactly solvable and, as in the case of the honeycomb lattice model, has two phases, respectively having a gapped and a gapless Majorana fermion spectrum. Furthermore, due to the higher dimension, the flux excitations form loops and are fermionic in nature \cite{ManSur(14)}. 

The iridate compound $\beta$-$\mathrm{Li_{2}IrO_{3}}$ has been found to be an approximate realization of the hyperhoneycomb lattice Kitaev model \cite{TakKat(15),KatYad(16)}. $\gamma$-$\mathrm{Li_{2}IrO_{3}}$ is another three-dimensional realization of the Kitaev model, on the stripy-honeycomb lattice \cite{ModSmi(14), BifJoh(14)}.

To probe the excitations (Majorana fermions and the flux loops), we can explore the dynamical response of the system. Previously, this has been done for the hyperhoneycomb lattice Kitaev model by calculating the dynamical structure factor \cite{SmiKno(16)} and by examining the defect generation during a linear quench \cite{SarRan(20)}. In this paper, we consider a different dynamical aspect of the model: how it responds when subjected to a periodic drive.   

Certain integrable many-particle systems, when subjected to periodic driving, show an unusual behaviour when the driving frequency is varied \cite{Das(10),HalDas(17),HalDas(22)}. As driving frequency increases, such systems do not respond monotonically. In particular, for certain combinations of driving frequency and amplitude, the system becomes locked in its initial state. This phenomenon, referred to as dynamical freezing, contradicts classical intuition. Classically, one would expect that as the driving frequency increases, the system's response would diminish and eventually vanish when the driving period becomes significantly shorter than the characteristic time scales within the system, with the system then remaining in its initial state for the entire duration of driving.

Dynamical freezing in an integrable many-particle system was first observed in the one-dimensional transverse-field Ising model \cite{Das(10)}.  In this model, when the magnetic field is harmonically driven at a fixed amplitude, the long-term average of the magnetization shows a distinct pattern of peaks and valleys as a function of the frequency, with the peaks indicating near-total saturation of the magnetization. Moreover, similar behavior was observed when periodic quenches replaced harmonic driving \cite{BhaDas(12)}.

Several aspects of dynamical freezing such as the effect of disorder \cite{RoyDas(15)}, the emergence of slow solitary oscillations \cite{BhaDas(12)}, the effect of interactions \cite{HalSen(21)}, and the ability to switch the freezing on and off by tuning parameters in the Hamiltonian \cite{DasMoe(12)} have been investigated in recent years. Notably, dynamical freezing has been demonstrated experimentally in a driven Ising chain \cite{HegKat(14)}. 

It is worth mentioning that dynamical freezing can be considered as a many-particle counterpart of single-particle phenomena studied in the past, such as the dynamical localization of a particle moving on a lattice in the presence of an alternating electric field \cite{DunKen(86), EckHol(09)}, and the  coherent destruction of tunneling of a particle moving in a periodically driven double-well potential \cite{GroDit(91),GroHan(92)}.

The dynamics of the 1D transverse-field Ising model, which has been extensively studied in the context of dynamical freezing, can be reduced to a two-level problem for each momentum value. However, dynamical freezing goes beyond two-level systems and has been demonstrated to occur in a bilayer graphene system, which features four energy bands \cite{SasSur(23)}. The three-dimensional Kitaev model, whose dynamics we investigate in this paper, is also a four-band system.

To make the Kitaev Hamiltonian periodic in time, we harmonically drive one of the spin-couplings as follows: $J_z(t) = J' + J \cos \omega t$, where $J'$ is a constant that contributes a nonzero average to $J_z(t)$. For nonzero $J'$, we observe dynamical freezing in the system.  When driven at a fixed amplitude $J$, the system shows a nonmonotonic response to changes in the driving frequency $\omega$, and freezes almost completely at some particular frequencies. The bias parameter $J'$ acts as a switch for the freezing---when $J'=0$, there is no freezing. Our results follow from numerically solving the time-dependent Schr\"odinger equation and analytical calculations based on a rotating wave approximation.
 
The rest of the paper is organized as follows. In Sec. \ref{s-kitaev}, we briefly review the 3D Kitaev model defined on a hyperhoneycomb lattice. In Sec. \ref{s-driving}, we present our results for the model subjected to a periodic drive, and we conclude with a discussion in Sec. \ref{s-summ}.

\section{\label{s-kitaev} Three-dimensional Kitaev model}


\begin{figure}
    \begin{center}
      \includegraphics[scale=1]{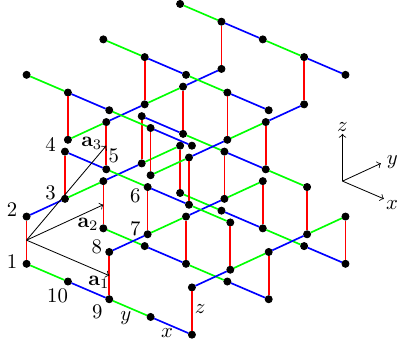}
    \caption{The hyperhoneycomb lattice: the four sites labeled
      $1,~2,~3,~4$ belong to a unit cell;
      $\textbf{a}_{1} = 2 \hat x$,~$\textbf{a}_{2} = 2 \hat y$, and $\textbf{a}_{3} = \hat x + \hat y + 2\hat z$ are the
      basis vectors; $x$ (blue), $y$ (green) and $z$ (red) are the three link types.}
    \label{f-lattice}
    \end{center}
    \end{figure}

The Kitaev Hamiltonian on the  hyperhoneycomb lattice \cite{ManSur(09)} has the same form as the original honeycomb lattice model \cite{Kit(06)}:
\begin{align}
\mathcal{H} &=-J_{x}\sum_{<j,k>_x} \sigma_j^x
\sigma_k^x - J_{y}\sum_{<j,k>_y} \sigma_j^y
\sigma_k^y - J_{z}\sum_{<j,k>_z} \sigma_j^z \sigma_k^z,
\label{e-kitham}
\end{align}
where $<j,k>_\alpha$ denotes a link labeled $\alpha$. The links are labeled as $x$, $y$ or $z$ as shown in Fig. \ref{f-lattice}. (The lattice shown in the figure has the same connectivity as the hyperhoneycomb lattice, which has a zig-zag structure for the chains.) Then the component of the Ising-like interaction along each link is determined by the corresponding link label. 

The elementary loops (called the plaquettes) in the hyperhoneycomb lattice contain ten sites (see Fig. \ref{f-lattice}). The Hamiltonian commutes with the $Z_2$ flux operators $W_p$ defined on the plaquettes as follows:
\begin{align}
W_p &= \prod_{j\in p} \sigma_j^{\beta_j},  
\label{e-flux}
\end{align}
where $j$ runs over the ten sites belonging to the plaquette $p$, and $\beta_j$ is the label for the link going out of the plaquette $p$ at the site $j$.

The above Hamiltonian can be solved exactly \cite{ManSur(09)} using a Majorana fermion representation of spin-1/2 \cite{Kit(06)}. Kitaev introduced four species of Majorana fermions at each site: $c_j, b_j^x, b_j^y$ and $b_j^z$. Then, the operators defined as
\begin{align}
    \tilde \sigma_j^\alpha = i b_j^\alpha c_j
    \label{e-majrep}
\end{align}
satisfy the spin-1/2 algebra when projected to the subspace corresponding to $b_j^x b_j^y b_j^z c_j = 1$. Writing the Hamiltonian in terms of the Majorana operators, it is straightforward to show that the operators $u_{jk}=ib_j^\alpha b_k^\alpha$, defined on the link connecting sites $j$ and $k$ (with $\alpha$ same as the link label), commute with $H$ as well as among themselves. In the Majorana representation, the plaquette operator becomes
\begin{align}
    W_p &= \prod_{<j,k>\in p} u_{jk},
    \label{e-flux-majorana}
\end{align}
where $<j,k>$ runs over the links in $p$.

It has been shown that for the ground state $W_p=1$ for all $p$. This corresponds to (for a particular gauge choice) $u_{jk} =1$ for all links $<j,k>$. Then, in terms of the Fourier modes of the Majorana fermions (which are standard complex fermions), the Hamiltonian becomes
\begin{align*}
    \mathcal{H} &= \sum_{\bf k}H({\bf k}),
\end{align*}
where,
\begin{align}
  H({\bf k}) & =\sum_{\textbf{k}}\Bigg[\frac{i}{2}\left\{e^{i k_{3}}\delta_1c_{1}^{\dagger}(\textbf{k})c_{4}(\textbf{k})+  \delta_2c_{3}^{\dagger}(\textbf{k})c_{2}(\textbf{k})\right\} \nonumber\\
  &\qquad + \dfrac{i}{2}\left\{ J_{z} c_{3}^{\dagger}(\textbf{k})c_{4}(\textbf{k})+J_{z} c_{1}^{\dagger}(\textbf{k})c_{2}(\textbf{k})\right\}+ h.c.\Bigg],
  \label{e-kham}
  \end{align} 
 where the subscript on the fermion operators labels the site index within a unit cell (see Fig. \ref{f-lattice}). Further,
\begin{align}
  \textbf{k}=k_{1}\thinspace\textbf{b}_{1}+
k_{2}\thinspace\textbf{b}_{2}+k_{3}\thinspace\textbf{b}_{3},  
\end{align}
with $k_i = 2\pi m_i,~m_i\in \mathbb{Z}$, and
\begin{align}
    \delta_j=J_{x}+J_{y}e^{-i k_{j}}, \mathrm{~for}~j=1,2.
\end{align}
Here
\begin{align}
    \textbf{b}_{1}=\dfrac{2\hat{\thinspace{{x}}}-\hat{z}}{4},~
\textbf{b}_{2}=\dfrac{2\thinspace\hat{y}-\hat{z}}{4},\thinspace \textbf{b}_{3}=\dfrac{\hat{z}}{2}
\end{align}
form a basis for the reciprocal lattice.  The Hamiltonian in Eq. (\ref{e-kham}) can be easily diagonalized to obtain the energy dispersions \cite{ManSur(09)}.

\section{\label{s-driving} Periodic driving}

We drive $J_z$ periodically, keeping $J_x=J_y=1$. Further, we focus on the regime where the driving amplitude is large compared to $J_x$ and $J_y$, and choose the initial state as the initial Hamiltonian's ground state. 

The Hamiltonian in Eq. (\ref{e-kham}) conserves the fermion number. In the ground state, the two negative energy lower bands are filled. Thus, if the initial state is the ground state at $t=0$, the dynamics for a given ${\bf k}$ will be restricted to the corresponding six-dimensional two-particle sector. 

To study the response to the driving, we define the following quantities \cite{SasSur(23)}:
\begin{align}
   q({\bf k}, t) &= \vert \bra{\psi_{{\bf k}}(0)} \ket{\psi_{{\bf k}}(t)}\vert^{2}, \label{e-qkt}\\ 
   \tilde{q}(t) &= \dfrac{1}{N}\sum_{\bf k} q({\bf k}, t), \label{e-qtilde}\\
   \bar q({\bf k}) &= \frac{1}{T} \int_0^T q({\bf k}, t)~ dt, \label{e-qbar}\\
    Q &= \frac{1}{NT} \sum_{\bf k} \int_0^T dt~ q({\bf k}, t) \nonumber\\
    &=  \frac{1}{T} \int_0^T dt~ \tilde{q}(t) = \frac{1}{N}\sum_{\bf k} \bar q({\bf k}), 
    \label{e-qavg}
    \end{align}
where $\ket{\psi_{\bf k}(0)}$ is the ground state of $H({\bf k})$ at $t=0$ and $N$ is the number of unit cells. The quantity $Q$ gives a measure of the extent of freezing; it is obtained by calculating the probability of the state with momentum ${\bf k}$ to remain in the initial state at a later time $t$ and then averaging over both ${\bf k}$ and $t$, where the time-average is taken over a long duration $T$. The maximum value of $Q$ is one, which would imply that the freezing is absolute and the system is stuck in the initial state throughout the evolution. 
\begin{figure*}
\includegraphics[width=\textwidth]{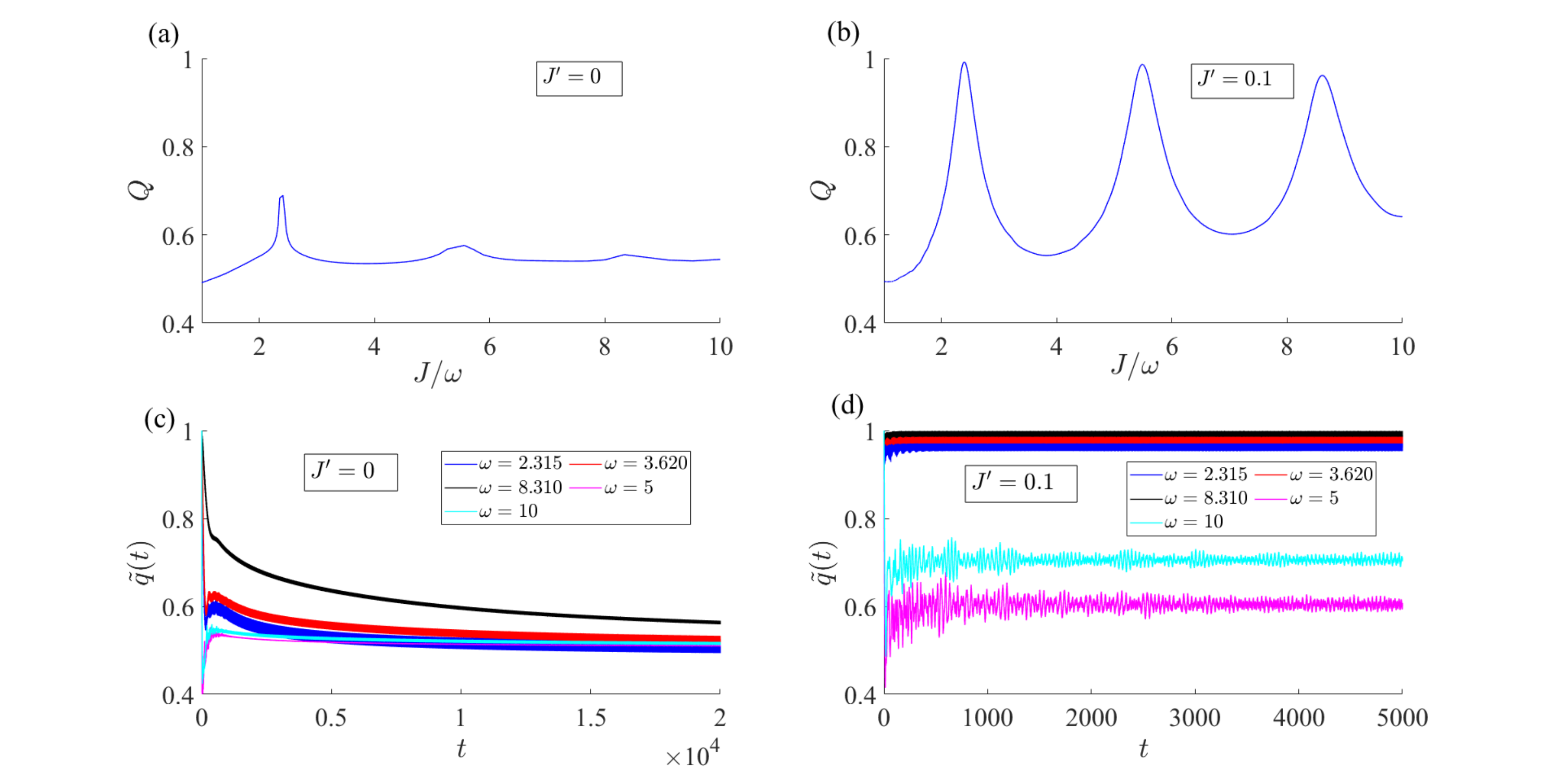}
\caption{\label{f-2p} $Q(J/\omega)$ for (a)~$J'=0 ~(N=3375)$, (b)~$J' = 0.1~ (N=1000)$; and $\tilde q(t)$ for various values of $\omega$ for (c)~$J'=0 ~(N=8000)$, (d)~$J' = 0.1~ (N=1000)$. In all the cases $J=20$.}
\end{figure*}

Here we note that the driving of $J_z$ does not violate the conservation of $W_p$ in the original spin model, and therefore no flux excitations are created due to the driving. Moreover, the ground state remains in the $W_p = +1$ sector throughout the driving, even though $J_z$ changes sign. To see this, we first observe that the hyperhoneycomb lattice consists of layers of chains on the $x$-$y$ plane containing only $J_x$ and $J_y$ interactions, and the chains lying on adjacent planes are then coupled via the $J_z$-interaction. Then, under the unitary transformation $\sigma_j^x \rightarrow \sigma_j^x$, $\sigma_j^y \rightarrow -\sigma_j^y$, $\sigma_j^z \rightarrow -\sigma_j^z$ of all spins lying on alternate $x$-$y$ planes, $J_x \rightarrow J_x$, $J_y \rightarrow J_y$, and $J_z \rightarrow -J_z$. Since each $W_p$ contains an even number of $\sigma_j^x$ and $\sigma_j^z$ operators undergoing the above transformation, it is invariant.

\subsection{Numerical analysis}

In an appropriately chosen basis, the two-particle Hamiltonian is given by (see Appendix \ref{a-2part-ham} for the derivation):
\begin{align}
 H_{2p}({\bf k}) &=
    \begin{bmatrix}
       -J_z & \beta^*_{\bf k} & 0 & 0 & -\beta_{\bf k} & 0\\\\
             \beta_{\bf k} & 0  & -\alpha_{\bf k} & \alpha_{\bf k} & 0 & -\beta_{\bf k} \\\\
       0 & -\alpha^*_{\bf k} & 0 & 0 & \alpha_{\bf k} & 0 \\\\
       0 & \alpha^*_{\bf k} & 0 & 0 & -\alpha_{\bf k} & 0 \\\\
       -\beta^*_{\bf k} & 0  & \alpha^*_{\bf k} & -\alpha^*_{\bf k} & 0 & \beta^*_{\bf k} \\\\
       0 & -\beta^*_{\bf k} & 0 & 0 & \beta_{\bf k} & J_z\\
    \end{bmatrix},
    \label{e-2pham}
    \end{align}
    where
\begin{align}    
    \alpha_{\bf k} &= \dfrac{ -e^{ik_{3}}(1+e^{-ik_{1}})- (1+e^{ik_{2}})}{4}, \label{e-alpha} \\ 
    \beta_{\bf k} &= \dfrac{ ie^{ik_{3}} (1+e^{-ik_{1}})-i (1+e^{ik_{2}})}{4}. \label{e-beta}
    \end{align}

We now drive the system by making $J_z$ time-dependent: $J_z(t) = J \cos{\omega t}$. For large $J$, during one cycle the Hamiltonian starts from the gapped phase ($J_z\gg J_x,~J_y$) and is driven through the gapless phase. Our initial state is the ground state at $t=0$, which, for $J\gg 1$ is the state $[1~ 0~ 0~ 0~ 0~ 0]$. We numerically compute $q({\bf k},t)$, the probability of remaining in the initial state at time $t$, for various values of driving frequency $\omega$, keeping $J$ fixed.

In Fig. \ref{f-2p}a we plot $Q$ ($q({\bf k}, t)$ averaged over both ${\bf k}$ and $t$) as a function of the dimensionless parameter $J/\omega$. (Throughout this paper, we use units in which $\hbar=1$.) Even though $Q$ varies nonmonotonically, showing some spikes, there is no freezing for any value of $J/\omega$. In Fig.  \ref{f-2p}c, we have plotted
$\tilde q(t)$ ($q({\bf k}, t)$ averaged over ${\bf k}$) for five different values of $\omega$. Three frequencies correspond to the peaks of $Q(J/\omega)$, represented by the graph's red, blue, and black curves. At these frequencies, there is a tendency to freeze at short time scales, but eventually, at larger times, the system oscillates among the initial ground state and the excited states, bringing down the value of $Q$.  

The absence of freezing with zero bias also occurs in the case of bilayer graphene, where the driving parameter is a chemical potential \cite{SasSur(23)}. In that case, the lack of freezing is shown to be related to a ground state degeneracy in a frame rotating with the driving. This prompts us to analyze our problem in an appropriate rotating frame. Moreover, we can then apply the method of rotating wave approximation in the high-frequency limit and obtain analytic solutions.   

\subsection{\label{s-rwa} Rotating wave approximation}

In the Rotating Wave Approximation (RWA) \cite{AshJoh(07), SenSen(21)}, we work in the interaction picture by transforming to a ``rotating" frame and then take the high-frequency limit. Then the Hamiltonian becomes (see Appendix \ref{s-app} for the details of the calculations in this section)
\begin{align}
 H'_{2p}({\bf k}) &=
    \begin{bmatrix}
       0 & \tilde\beta^*_{\bf k} & 0 & 0 & -\tilde\beta_{\bf k} & 0\\\\
             \tilde\beta_{\bf k} & 0  & -\alpha_{\bf k} & \alpha_{\bf k} & 0 & -\tilde\beta_{\bf k} \\\\
       0 & -\alpha^*_{\bf k} & 0& 0 & \alpha_{\bf k} & 0 \\\\
       0 & \alpha^*_{\bf k} & 0 & 0 & -\alpha_{\bf k} & 0 \\\\
       -\tilde\beta^*_{\bf k} & 0  & \alpha^*_{\bf k} & -\alpha^*_{\bf k} & 0 & \tilde\beta^*_{\bf k} \\\\
       0 & -\tilde\beta^*_{\bf k} & 0 & 0 & \tilde\beta_{\bf k} & 0\\
    \end{bmatrix}.
    \label{e-rwaham-nobias}
    \end{align}
where $\tilde\beta_{\bf k} =\beta_{\bf k} J_0\left(J/\omega\right) $, with $J_0\left(J/\omega\right)$ being the zeroth order Bessel's function of the first kind. As shown in Appendix \ref{s-app}, we find that for all values of ${\bf k}$ and $\omega$, $Q$ satisfies the following bound in the thermodynamic limit:
\begin{align}
    \frac{5}{16} \le Q \le \frac{3}{8}.
    \label{e-Qbound2p}
\end{align}
Even though we have not explicitly calculated $Q$, our RWA analysis tightly bounds $Q$ between 0.3125 and 0.375. However, the numerically computed values of $Q$ (Fig. \ref{f-2p}a)  are above the RWA upper bound for all values of $\omega$. As we argue below, this is not a failure of our RWA calculations, where the time averaging is done in the infinite-time limit, but a manifestation of the finite time over which the numerical averages are calculated.

\subsubsection{\label{s-slowmodes} Effect of slow modes}

The RWA expression for $\bar q({\bf k})$ in Eq. (\ref{e-qbarrw}) is valid only in the limit $T\rightarrow \infty$ (where $T$ is the time over which the averaging is done). For finite $T$, the infinite-$T$ limit is a good approximation provided $T^{-1}$ is small compared to $\mu_3,~ \mu_5$, and $(\mu_5-\mu_3)$, the frequencies contained in $q({\bf k}, t)$ [Eq. (\ref{e-qktrw})].
As $\mu_3,~ \mu_5$, or $(\mu_5-\mu_3)$ approaches zero, the $T$-value required for a good agreement between numerical analysis and  RWA diverges. 

As shown in Appendix \ref{s-app}, when any among $\mu_3,~ \mu_5$, or $(\mu_5-\mu_3)$ approaches zero, the value of $\bar q({\bf k})$ lies above the generic case upper bound of 3/8. The number of modes for which this occurs is $O(N^{2/3})$, provided $J_0(J/\omega) \neq 0$. This explains why for finite $T$ and $N$, $Q(\omega)$ lies above the upper bound of 3/8 obtained in the limit $N,~T\rightarrow \infty$ (Fig. \ref{f-2p}a). When $J_0(J/\omega) = 0$, $\bar q({\bf k})$ overshoots 3/8 for all values of ${\bf k}$. This accounts for the spikes in $Q(\omega)$, which appear around $\omega$-values for which $J_0(J/\omega)$ vanishes.

\begin{figure*}
   \includegraphics[width=\textwidth]{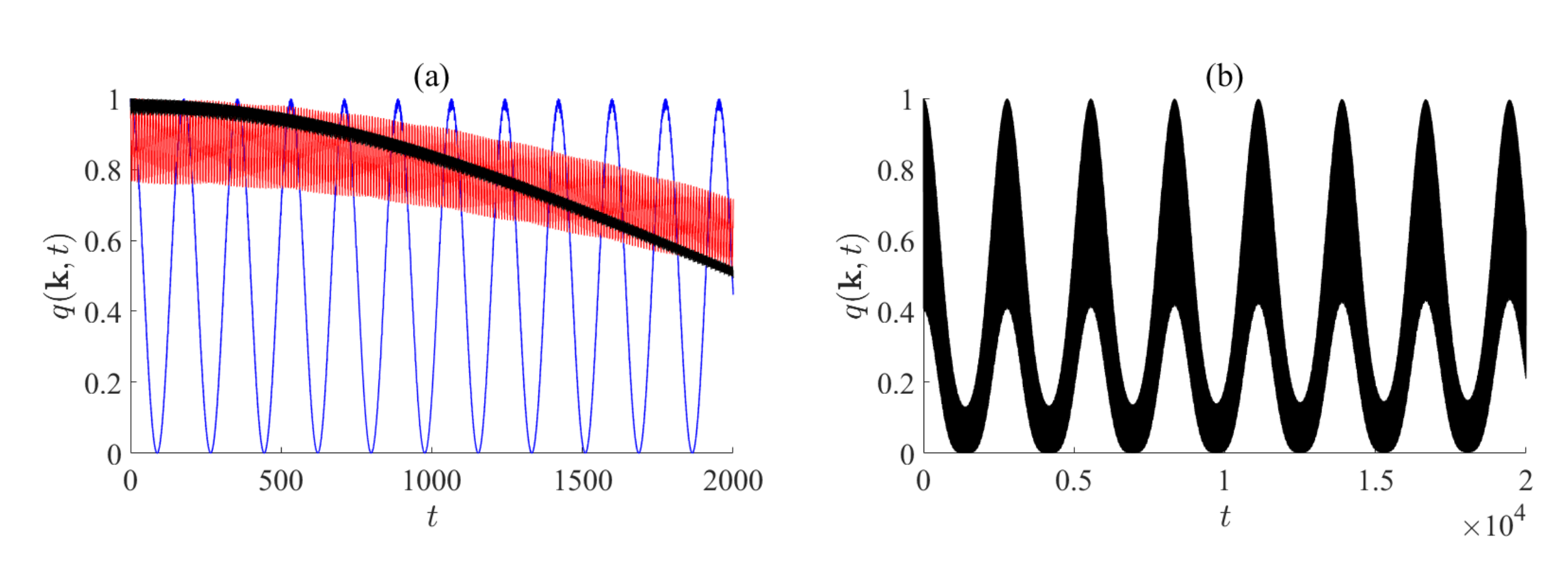}
  \caption{\label{f-singlemode} (a) $q({\bf k}, t)$ for three different values of ${\bf k}$ for $J'=0$, $J=20$: $(k_1, k_2, k_3) = (\pi/8,-\pi/3,\pi/6)$  and $J/\omega = 6.667$ (blue); $(k_1, k_2, k_3) = (\pi/8,-\pi/3,\pi/6)$  and $J/\omega = 8.628$ (red); $ (k_1, k_2, k_3) = (\pi/4.1,\pi/4,0)$ and $J/\omega = 6.667$ (black), and (b) $q({\bf k}, t)$ plotted for a longer duration for the third case (black).}
  \end{figure*}
  
In Fig. \ref{f-singlemode}a, we have plotted $q({\bf k},t)$ for three representative values of ${\bf k}$. For a generic value of ${\bf k}$ and $J/\omega$ (blue curve), $q({\bf k}, t)$ oscillates many times during the evolution and its time-average $\bar q({\bf k}) = 0.3733$ compares well with the RWA value of $0.3711$ obtained from Eq. (\ref{e-qbarrw}). The dynamics slows down when either $J_0(J/\omega) \approx 0$ or $\alpha_{\bf k} \beta^*_{\bf k} \approx \alpha^*_{\bf k} \beta_{\bf k}$. Red and black curves, respectively, show the slowing down for specific instances of these two cases. In the duration shown, $q({\bf k}, t)$ is well above the RWA value.  
Fig. \ref{f-singlemode}b plots $q({\bf k},t)$ corresponding to the third ${\bf k}$ value (black curve) for a longer time. Then the time-average $\bar q({\bf k})$ becomes 0.3371 while the RWA formula yields {0.3750}.

\subsection{Driving with bias}

We have found that when $J_z$ is driven as \mbox{$J_z(t) = J\cos{\omega t}$}, the system does not freeze under any driving condition, even though the dynamics slows down for certain combinations of $J$ and $\omega$. When $J_0\left(J/\omega\right) = 0$, the initial state becomes an eigenstate of the rotating wave Hamiltonian for all values of ${\bf k}$, which would indicate that the system freezes. However, at the same time the corresponding eigenvalue also becomes degenerate, which restores the dynamics, as we have described in Appendix \ref{s-app}.

 The degeneracy in the Hamiltonian can be lifted by adding a bias $J'$ to the driving \cite{SasSur(23)}: $J_z(t) = J' + J\cos{\omega t}$. The rotating wave Hamiltonian in the two-particle sector then becomes
\begin{align}
 H'_{2p}({\bf k}) &=
    \begin{bmatrix}
       0 & \tilde\beta^*_{\bf k} & 0 & 0 & -\tilde\beta_{\bf k} & 0\\\\
             \tilde\beta_{\bf k} & J'  & -\alpha_{\bf k} & \alpha_{\bf k} & 0 & -\tilde\beta_{\bf k} \\\\
       0 & -\alpha^*_{\bf k} & J'& 0 & \alpha_{\bf k} & 0 \\\\
       0 & \alpha^*_{\bf k} & 0 & J' & -\alpha_{\bf k} & 0 \\\\
       -\tilde\beta^*_{\bf k} & 0  & \alpha^*_{\bf k} & -\alpha^*_{\bf k} & J' & \tilde\beta^*_{\bf k} \\\\
       0 & -\tilde\beta^*_{\bf k} & 0 & 0 & \tilde\beta_{\bf k} & 2J'\\
    \end{bmatrix},
    \label{e-rwaham-bias}
    \end{align}
where, as defined earlier,  $\tilde\beta_{\bf k} = J_0(J/\omega) \beta_{\bf k}$, and $\alpha_{\bf k}$ and $\beta_{\bf k}$ are given in Eqs. (\ref{e-alpha}) and (\ref{e-beta}). When $J_0(J/\omega)=0$, the initial state is still an eigenstate of $H'_{2p}({\bf k})$, but, unlike in the bias-free case, this state no longer has a degenerate partner due to the presence of $J'$. Therefore, $q({\bf k},t) = 1$ for all ${\bf k}$, and thus $Q=1$.

We have plotted $Q(J/\omega)$ in Fig. \ref{f-2p}b. The peak values of $Q$ are close to one, showing that the system is almost completely frozen at those frequencies.  The peaks occur at $J/\omega = 2.4,~5.5,$ and $8.6$. This compares well with the zeros of $J_0(x)$, which are at $x =$ 2.405, 5.520, and 8.654.

In Fig. \ref{f-2p}d we have shown $\tilde q(t)$ for five different values of $\omega$. At the three frequencies that correspond to the peaks in $Q(J/\omega)$ (red, blue, and black curves in the figure), $\tilde q(t)$  remains close to one at all times.

\begin{figure}
    \includegraphics[width=.46\textwidth]{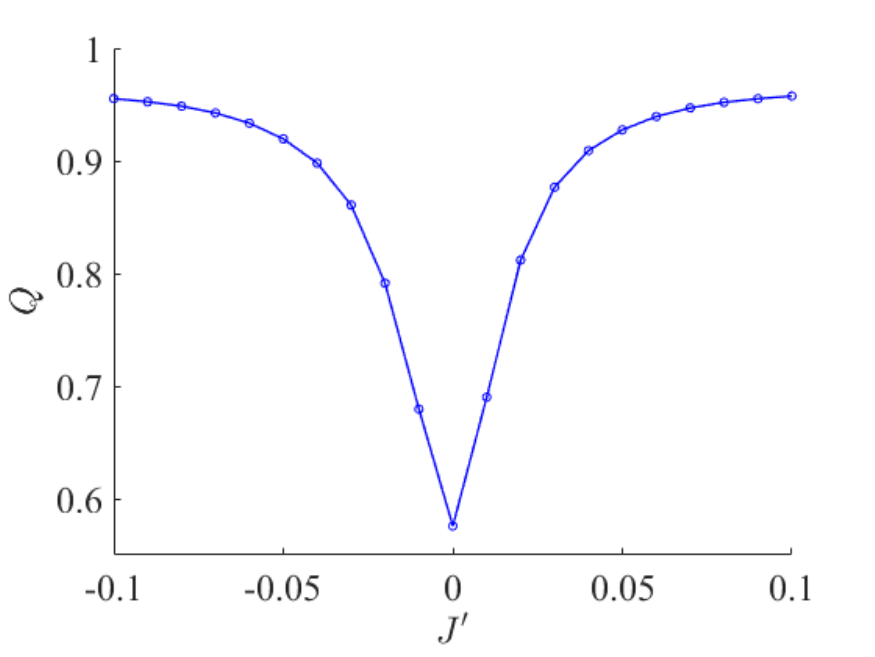}
\caption{$Q(J')$ for $N=8000$, $J=20$ and $\omega = 2.320$.}
\label{f-switch}
\end{figure}

\section{\label{s-summ} Summary and discussion}

In this paper, we have studied the response of the three-dimensional Kitaev model on a hyperhoneycomb lattice to strong and fast driving of one of the spin-couplings. When the driving has no bias, there is no dynamical freezing; the system oscillates among the initial ground state and other excited states, and, for a given driving amplitude, its response is nearly independent of the driving frequency. Our calculations based on a rotating wave approximation support these observations. The rotating wave analysis further shows that dynamical freezing is switched on by introducing a bias to the driving. In the presence of a bias, the system freezes almost absolutely for certain ratios between the amplitude and frequency of the driving. These ratios match well with the values predicted by the rotating wave analysis.

The switching of dynamical freezing is illustrated in Fig. \ref{f-switch}, which plots the degree of freezing $Q$ against the bias parameter $J'$, for a value of frequency at which system freezes for nonzero $J'$. Such switching of dynamical freezing has been previously observed in the transverse-field $X$-$Y$ model \cite{DasMoe(12)} and in a tight-binding model of bilayer graphene. We also note that the two-dimensional Kitaev model undergoes dynamical freezing even under bias-free driving and, therefore, exhibits no switching. This is because the one-particle Hamiltonian for the 2D model corresponds to a two-level system and is analogous to the transverse-field Ising model.  

We conclude with a discussion of the experimental implications of our model. As mentioned in the introduction, $\beta$-$\mathrm{Li_{2}IrO_{3}}$ is considered to be a good candidate to realize the hyperhoneycomb lattice Kitaev Hamiltonian. However, for the driven Hamiltonian, a more plausible realization is via cold atoms on an optical lattice. For the honeycomb lattice Kitaev Hamiltonian, Duan, et. al. \cite{DuaDem(03)} have proposed a technique that exploits spin-dependent tunneling between neighboring atoms in an optical lattice to induce the anisotropic spin interactions. 
This method could in principle be extended to realize the three-dimensional model as well. The tunneling amplitudes in an optical lattice can be made time-periodic by introducing an oscillating frequency difference between the two counter-propagating  lasers that create the lattice potential in a given direction \cite{LigSia(07)}. In general, this will make all parameters in the effective spin-Hamiltonian time dependent. In this paper, we have considered the simplest scenario in which only one of the couplings ($J_z$) is driven while keeping the other two ($J_x$ and $J_y$) fixed. Modeling the periodically driven optical lattice system will require the implementation of more complex driving protocols.

\appendix

\section{\label{a-2part-ham} Two-particle Hamiltonian}

To compute the two-particle Hamiltonian, first, we write down the matrix representation of the one-particle Hamiltonian. For each ${\bf k}$-value, we label the basis states using the occupation number $n_i({\bf k})$ for each sublattice: $ \ket{n_1({\bf k}), n_2({\bf k}), n_3({\bf k}), n_4({\bf k})}$.
Then the four basis states in the one-particle sector are 
\begin{align*}
  \ket{1}_{1p} = \ket{ 1 0 0 0}, & ~~~~~\ket{2}_{1p} =\ket{ 0 1 0 0},\\
  \ket{3}_{1p} =\ket{ 0 0 1 0}, & ~~~~~\ket{4}_{1p} =\ket{ 0 0 0 1}.  
\end{align*}
Then,
 \begin{align*}
 H_{1p}({\bf k})=
 \begin{bmatrix}
     0 &\dfrac{i}{2} J_{z} & 0 & \dfrac{i}{2}e^{ik_{3}}\delta_1\\\\
  \dfrac{-i}{2} J_{z} & 0 &\dfrac{-i}{2} \delta_2^{*} & 0\\\\
    0 &\dfrac{i}{2} \delta_2 & 0 & \dfrac{i}{2}J_{z}  \\\\
    -\dfrac{i}{2}e^{-i k_{3}}\delta_1^{*} & 0 &\dfrac{-i}{2} J_{z} & 0 \\ 
 \end{bmatrix}.
 \label{e-opham}
 \end{align*}
After employing the unitary transformation 
\begin{align*}
U=\dfrac{1}{\sqrt{2}}
\begin{bmatrix}
    1 & -i & 0 & 0\\     
    0 & 0 & 1 & -i\\
    -i & 1 & 0 & 0 \\
    0 & 0 & -i & 1 
\end{bmatrix},
\end{align*}
the time-dependent $J_z$-terms in the Hamiltonian become diagonal:
\begin{align}
H_{1p}({\bf k}) &=
\begin{bmatrix}
 -\dfrac{ J_{z}}{2} &\alpha_{\bf k}& 0 & \beta_{\bf k} \\
  \alpha_{\bf k}^{*} & -\dfrac{J_{z}}{2} & \beta_{\bf k}^{*} & 0\\
    0 & \beta_{\bf k} & \dfrac{J_{z}}{2} & -\alpha_{\bf k}\\
    \beta_{\bf k}^{*} & 0 & -\alpha_{\bf k}^{*} & \dfrac{J_{z}}{2}
\end{bmatrix},
\end{align} 
where
\begin{align}    
    \alpha_{\bf k} &= \dfrac{ -e^{ik_{3}}(1+e^{-ik_{1}})- (1+e^{ik_{2}})}{4}, \label{ea-alpha} \\ 
    \beta_{\bf k} &= \dfrac{ ie^{ik_{3}} (1+e^{-ik_{1}})-i (1+e^{ik_{2}})}{4}. \label{ea-beta}
    \end{align}
Choosing the two-particle basis states as follows,
\begin{align}
\begin{array}{lcl}
    \ket{1}_{2p} &=& \frac{1}{\sqrt{2}} \left( \ket{1}_{1p} \otimes \ket{2}_{1p} - \ket{2}_{1p} \otimes \ket{1}_{1p} \right)\\
    \ket{2}_{2p} &=& \frac{1}{\sqrt{2}} \left( \ket{1}_{1p} \otimes \ket{3}_{1p} - \ket{3}_{1p} \otimes \ket{1}_{1p} \right)\\
    \ket{3}_{2p} &=& \frac{1}{\sqrt{2}} \left( \ket{1}_{1p} \otimes \ket{4}_{1p} - \ket{4}_{1p} \otimes \ket{1}_{1p} \right)\\
    \ket{4}_{2p} &=& \frac{1}{\sqrt{2}} \left( \ket{2}_{1p} \otimes \ket{3}_{1p} - \ket{3}_{1p} \otimes \ket{2}_{1p} \right)\\
    \ket{5}_{2p} &=& \frac{1}{\sqrt{2}} \left( \ket{2}_{1p} \otimes \ket{4}_{1p} - \ket{4}_{1p} \otimes \ket{2}_{1p} \right)\\
    \ket{6}_{2p} &=& \frac{1}{\sqrt{2}} \left( \ket{3}_{1p} \otimes \ket{4}_{1p} - \ket{4}_{1p} \otimes \ket{3}_{1p} \right),
    \end{array}
    \label{e-2pbasis}
    \end{align}
we obtain the following two-particle Hamiltonian:
\begin{align}
 H_{2p}({\bf k}) &=
    \begin{bmatrix}
       -J_z & \beta^*_{\bf k} & 0 & 0 & -\beta_{\bf k} & 0\\\\
             \beta_{\bf k} & 0  & -\alpha_{\bf k} & \alpha_{\bf k} & 0 & -\beta_{\bf k} \\\\
       0 & -\alpha^*_{\bf k} & 0 & 0 & \alpha_{\bf k} & 0 \\\\
       0 & \alpha^*_{\bf k} & 0 & 0 & -\alpha_{\bf k} & 0 \\\\
       -\beta^*_{\bf k} & 0  & \alpha^*_{\bf k} & -\alpha^*_{\bf k} & 0 & \beta^*_{\bf k} \\\\
       0 & -\beta^*_{\bf k} & 0 & 0 & \beta_{\bf k} & J_z\\
    \end{bmatrix}.
    \label{ea-2pham}
    \end{align}

\section{\label{s-app} RWA calculation of $\bar q ({\bf k})$ for $J'=0$}

To implement the rotating wave approximation, we first transform to a rotating frame by doing the following time-dependent unitary transformation in the one-particle subspace:
\begin{align}
U &=\exp[-i\left(\frac{J}{2 \omega}\sin{(\omega t)}\right)(\sigma^{z}\otimes I)].
\label{e-uint1}
\end{align} 
Then, the effective one-particle Hamiltonian that governs the dynamics in the rotating frame, given by \mbox{$H_{1p}' = UH_{1p}U^\dagger + i \partial_t U U^\dagger$}, becomes
\begin{equation}
  H'_{1p}({\bf k})= 
\begin{bmatrix}
  0 & \alpha_{\bf k} & 0 & \beta_{\bf k}e^{-2i \theta}    \\\\
    \alpha_{\bf k}^{*} & 0 & \beta_{\bf k}^{*}e^{-2i \theta} & 0  \\\\
    0 & \beta_{\bf k}e^{2i \theta} & 0 & -\alpha_{\bf k}      \\\\
    \beta_{\bf k}^{*}e^{2i \theta} & 0 & -\alpha_{\bf k}^{*} & 0 \\
\end{bmatrix},
\label{e-hint}
\end{equation}
where $\theta=\left(J/2\omega\right)\sin{\omega t}$. Next we do an expansion of $\exp[i(J/\omega)\sin ({\omega t})]$ in terms of $e^{in\omega t}$, where $n\in \mathbb{Z}$:
\begin{align}
    \exp[i\left(\dfrac{J}{\omega}\right) \sin (\omega t)]  = \sum_{n=-\infty}^{\infty}  J_{n}\left(\dfrac{J}{\omega}\right) e^{i n \omega t},
    \label{e-bessel}
\end{align}
where $J_n(J/\omega)$ are the Bessel's functions of the first kind.

For large $\omega$, the predominant term in the above sum is when $n=0$. In the rotating wave approximation (RWA), we neglect all faster-oscillating terms, which correspond to $n>0$ \cite{AshJoh(07)}. Then, the Hamiltonian becomes
\begin{equation}
 H'_{1p}({\bf k})= 
\begin{bmatrix}
 0 & \alpha_{\bf k} & 0 & \tilde\beta_{\bf k}   \\\\
   \alpha_{\bf k}^{*} & 0 & \tilde\beta_{\bf k}^* & 0    \\\\ 
    0 & \tilde\beta_{\bf k} & 0 & -\alpha_{\bf k}    \\\\
    \tilde\beta_{\bf k}^*& 0 & -\alpha_{\bf k}^{*} & 0 \\
 \end{bmatrix},
 \label{e-1prwah}
\end{equation}
where $\tilde\beta_{\bf k} =\beta_{\bf k} J_0\left(J/\omega\right)$. The corresponding Hamiltonian in the two-particle sector is then
\begin{align}
 H'_{2p}({\bf k}) &=
    \begin{bmatrix}
       0 & \tilde\beta^*_{\bf k} & 0 & 0 & -\tilde\beta_{\bf k} & 0\\\\
             \tilde\beta_{\bf k} & 0  & -\alpha_{\bf k} & \alpha_{\bf k} & 0 & -\tilde\beta_{\bf k} \\\\
       0 & -\alpha^*_{\bf k} & 0& 0 & \alpha_{\bf k} & 0 \\\\
       0 & \alpha^*_{\bf k} & 0 & 0 & -\alpha_{\bf k} & 0 \\\\
       -\tilde\beta^*_{\bf k} & 0  & \alpha^*_{\bf k} & -\alpha^*_{\bf k} & 0 & \tilde\beta^*_{\bf k} \\\\
       0 & -\tilde\beta^*_{\bf k} & 0 & 0 & \tilde\beta_{\bf k} & 0\\
    \end{bmatrix}.
    \label{ea-rwaham-nobias}
    \end{align}

    The sublattice symmetry is made manifest by interchanging indices 2 and 6:
    \begin{align}
     H_{2p}'({\bf k}) &=
    \begin{bmatrix}
    0 & 0 & 0 & 0 & -\tilde\beta^*_{\bf k} & \tilde\beta_{\bf k} \\\\
    0 & 0 & 0 & 0 & \tilde\beta^*_{\bf k} & -\tilde\beta_{\bf k} \\\\
    0 & 0 & 0 & 0 & \alpha^*_{\bf k} & -\alpha_{\bf k} \\\\
    0 & 0 & 0 & 0 & -\alpha^*_{\bf k} & \alpha_{\bf k} \\\\
    -\tilde\beta_{\bf k} & \tilde\beta_{\bf k}  & \alpha_{\bf k} & -\alpha_{\bf k} & 0 & 0 \\\\
    \tilde\beta^*_{\bf k} & -\tilde\beta^*_{\bf k}  & -\alpha^*_{\bf k} & \alpha^*_{\bf k} & 0 & 0 \\
    \end{bmatrix}.
    \end{align}
The eigenvalues of $H_{2p}({\bf k})$ are 
\begin{align}
\label{ea-eigval}
\begin{split}
  &  \mu_1  = \mu_2  = 0, \\
 & \mu_3 = \sqrt{2\left(|\alpha_{\bf k}|^2 + |\tilde \beta_{\bf k}|^2  - |\alpha_{\bf k}^2 + \tilde \beta_{\bf k}^2| \right)},\\
 & \mu_4  = -\sqrt{2\left(|\alpha_{\bf k}|^2 + |\tilde \beta_{\bf k}|^2  - |\alpha_{\bf k}^2 + \tilde \beta_{\bf k}^2| \right)},\\
 &\mu_5  = \sqrt{2\left(|\alpha_{\bf k}|^2 + |\tilde \beta_{\bf k}|^2  + |\alpha_{\bf k}^2 + \tilde \beta_{\bf k}^2| \right)}, \\
 &\mu_6  =  -\sqrt{2\left(|\alpha_{\bf k}|^2 + |\tilde \beta_{\bf k}|^2  + |\alpha_{\bf k}^2 + \tilde \beta_{\bf k}^2| \right)},
 \end{split}
\end{align}
and the respective eigenvectors are,
\begin{align*}
    \ket{\mu_1}= \frac{1}{\sqrt{2}}
   \begin{bmatrix}
     1\\
     1\\
     0\\
     0\\
     0\\
     0
     \end{bmatrix},~
     \ket{\mu_2}=\frac{1}{\sqrt{2}}
    \begin{bmatrix}
     0\\
     0\\
     1\\
     1\\
     0\\
     0
     \end{bmatrix},
     \end{align*}
     \begin{align*}
     \ket{\mu_3} = \frac{1}{\mathcal{N}}
     \begin{bmatrix}
     -(\tilde{\beta}_{\bf k}-\tilde{\beta}^*_{\bf k}\eta)\\
     \tilde{\beta}_{\bf k}-\tilde{\beta}^*_{\bf k}\eta\\
     {\alpha}_{\bf k} - {\alpha}_{\bf k}^*\eta\\
     -({\alpha}_{\bf k} - {\alpha}_{\bf k}^*\eta)\\
    \mu_3\\
   \eta\mu_3
     \end{bmatrix},~
     \ket{\mu_4} = \frac{1}{\mathcal{N}}
     \begin{bmatrix}
     -(\tilde{\beta}_{\bf k}-\tilde{\beta}^*_{\bf k}\eta)\\
     \tilde{\beta}_{\bf k}-\tilde{\beta}^*_{\bf k}\eta\\
     {\alpha}_{\bf k} - {\alpha}^*_{\bf k}\eta\\
     -({\alpha}_{\bf k} - {\alpha}^*_{\bf k}\eta)\\
    -\mu_3\\
   -\eta\mu_3
     \end{bmatrix},
     \end{align*}
     \begin{align*}
     \ket{\mu_5} = \frac{1}{\mathcal{N}'}
     \begin{bmatrix}
     -(\tilde{\beta}_{\bf k}+\tilde{\beta}^*_{\bf k}\eta)\\
     \tilde{\beta}_{\bf k}+\tilde{\beta}^*_{\bf k}\eta\\
     {\alpha}_{\bf k} + {\alpha}^*_{\bf k}\eta\\
     -({\alpha}_{\bf k} + {\alpha}^*_{\bf k}\eta)\\
    \mu_5\\
   -\eta\mu_5
     \end{bmatrix},~
     \ket{\mu_6} = \frac{1}{\mathcal{N}'}
     \begin{bmatrix}
     -(\tilde{\beta}_{\bf k}+\tilde{\beta}^*_{\bf k}\eta)\\
     \tilde{\beta}_{\bf k}+\tilde{\beta}^*_{\bf k}\eta\\
     {\alpha}_{\bf k} + {\alpha}^*_{\bf k}\eta\\
     -({\alpha}_{\bf k} + {\alpha}^*_{\bf k}\eta)\\
    -\mu_5\\
   \eta\mu_5
     \end{bmatrix}
\end{align*}
where $\mathcal{N}$ and $\mathcal{N}'$ are normalization factors, and 
\begin{align*}
\eta=\dfrac{\vert  {\alpha}_{\bf k}^{2}+\tilde\beta_{\bf k}^{2}\vert} {{\alpha}^{*2}_{\bf k} + \tilde\beta^{*2}_{\bf k}}.
\end{align*}

Let $\ket{i}$ denote the canonical basis vectors $(0,\dots~ 1,\dots~ 0)$, where the $i$-th element is 1 and the rest zero, and let
\begin{align*}
    x_{i,j} = \frac{\bra{i}\ket{\mu_j}}
    {\bra{\mu_j}\ket{\mu_j}^{\frac{1}{2}}}.
\end{align*}
The initial state $\ket{\psi_{\bf k}(0)}=\ket{1}$, therefore, 
\begin{align*}
    {\bra{\psi_{\bf k}(0)} \ket{\psi_{\bf k}(t)}} &= \sum_j |x_{1,j}|^2 e^{-i\mu_j t} \\
    &= \frac{1}{2} + 2|x_{1,3}|^2 \cos {\mu_3 t} + 2|x_{1,5}|^2 \cos {\mu_5 t},
\end{align*}
since $|x_{1,4}|^2 = |x_{1,3}|^2$, $|x_{1,6}|^2 = |x_{1,5}|^2$, $\mu_4= -\mu_3$ , and $\mu_6= -\mu_5$.
Normalization of $\ket{\psi_{\bf k}(t)}$  implies that \begin{align}
    |x_{1,3}|^2 + |x_{1,5}|^2 = 1/4.
    \label{e-x31norm}
\end{align} 
Let $|x_{1,3}|^2 = u_{\bf k}$. Then,
\begin{align}
   {\bra{\psi_{\bf k}(0)}\ket{\psi_{\bf k}(t)}} = \frac{1}{2} + 2u_{\bf k} \cos {\mu_3 t} 
    + \left( \frac{1}{2} - 2 u_{\bf k} \right) \cos {\mu_5 t},
    \label{ea-psi0psit}
\end{align}
where
\begin{align}
     u_{\bf k} = \dfrac{2 \vert \tilde{\beta}_{\bf k}\vert^{2} \vert{{\alpha}_{\bf k}}^{2}+{\tilde\beta}_{\bf k}^{2}\vert - \left(2|\tilde{\beta}_{\bf k}|^{4} +   {{\alpha}^{*2}_{\bf k}} {\tilde{\beta}}_{\bf k}^{2} + {{\alpha}_{\bf k}}^{2} {{\tilde\beta}^{*2}_{\bf k}}\right)}
     {8\vert{{\alpha}_{\bf k}}^{2}+ {\tilde\beta}_{\bf k}^{2}\vert \left(\vert {\alpha}_{\bf k}\vert^{2}+\vert {\tilde\beta}_{\bf k}\vert^{2}- \vert {\alpha}_{\bf k}^{2}+ {\tilde\beta}_{\bf k}^{2}\vert\right)}. 
\end{align}
Evaluating $q({\bf k},t)$, we get
\begin{align}
    q({\bf k},t)&= \vert \bra{\psi_{{\bf k}}(0)} \ket{\psi_{{\bf k}}(t)}\vert^{2} \nonumber \\
    &= \Bigg[ \frac{1}{2} + 2 u_{\bf k} \cos {\mu_3 t} + \left( \frac{1}{2} - 2 u_{\bf k}\right) \cos {\mu_5 t}\Bigg]^2.
    \label{e-qktrw}
    \end{align}
   Next, we calculate $\bar q({\bf k})$, the time-average of $q({\bf k},t)$. $q({\bf k},t)$ contains four frequencies: $\mu_3, \mu_5, (\mu_3 + \mu_5)$, and $(\mu_3 - \mu_5)$. When none of these frequencies is zero, i.e., $\mu_3 \neq 0, \mu_5 \neq 0$, and $\mu_3 \neq \mu_5$, and as $T\rightarrow \infty$, we get
\begin{align}
     \bar q({\bf k}) &= \lim_{T\rightarrow \infty} \frac{1}{T} \int_0^T dt~ \left|{}_{2p}{\bra{\psi_{\bf k}(0)}\ket{\psi_{\bf k}(t)}}_{2p}\right|^{2}, \nonumber\\
     &= \dfrac{3}{8} + 4u_{\bf k}^2 - u_{\bf k}.
   \label{e-qbarrw}
\end{align}
Normalization of $\ket{\psi_{\bf k}(t)}_{2p}$ [Eq. (\ref{e-x31norm})] implies that
\begin{align}
    0 \le u_{\bf k} \le 1/4.
    \label{e-ubound}
\end{align}
Consequently,
\begin{align}
    \frac{5}{16} \le \bar q({\bf k}) \le \frac{3}{8}.
    \label{e-qbarbound2p}
\end{align}

 We now consider the cases where some frequency in $q({\bf k},t)$ is zero. There are three possibilities:  \mbox{$a)$ $\mu_3=\mu_5\neq 0$}, $b)$ $\mu_3=0$, and $c)$ $\mu_5=0$.

\paragraph{} $\mu_3=\mu_5\neq 0$ implies that either \mbox{$k_1=\pi$} or \mbox{$k_2=\pi$}, but not both. Then,
\begin{align*}
q({\bf k}, t) = \frac{1}{4}(1+ \cos \mu_3 t)^2,
\end{align*}
and
\begin{align*}
   \bar q({\bf k}) = \frac{3}{8}, 
\end{align*}
which coincides with the upper bound for the generic case given by Inequality (\ref{e-qbarbound2p}).

\paragraph{} $\mu_3 = 0$ implies that 
\begin{align}
  \alpha_{\bf k} \tilde\beta^*_{\bf k} = \alpha^*_{\bf k} \tilde\beta_{\bf k}.
  \label{e-spcase1}
  \end{align} 
The above condition can be satisfied in two ways: 1) $k_1=k_2$, or $k_1+k_2 = 2\pi$ for arbitrary $J$ and $\omega$, and 2) $J_0(J/\omega)=0$ (and, consequently, $\tilde \beta_{\bf k}=0$), for all values of ${\bf k}$. In the first case, 
\begin{align}
    q({\bf k},t)= \Bigg[ \frac{1}{2} + 2 u_{\bf k} + \left( \frac{1}{2} - 2 u_{\bf k}\right) \cos {\mu_5 t}\Bigg]^2,
    \end{align}
and 
\begin{align}
\bar q({\bf k})=\dfrac{3}{8} + u_{\bf k} + 6u_{\bf k}^2.
\end{align}
From the bound on $u_{\bf k}$ given by the Inequality (\ref{e-ubound}), it immediately follows that $3/8 \le \bar q({\bf k}) \le 1$, which lies above the upper bound in the generic case. However, since $k_1=k_2$ or $k_1+k_2 = 2\pi$ when $\mu_3=0$, the number of modes satisfying this condition is $O(N^{\frac{2}{3}})$, therefore, their contribution to $Q$ can be neglected in the thermodynamic limit. 

In the second scenario, Eq. (\ref{e-spcase1}) is satisfied via $J_0(J/\omega)=0$ independent of ${\bf k}$. Then, the initial state $[1~0 ~0~ 0~ 0~ 0]$ becomes an eigenstate of the $H'_{2p}({\bf k})$ [Eq. (\ref{ea-rwaham-nobias})] for all ${\bf k}$, which would have implied that the state is stationary and therefore $\bar{q}({\bf k}) = 1$. However, when $\tilde\beta_{\bf k} = 0$, the initial state becomes degenerate with the state $[0~ 0~ 0~ 0~ 0~ 1]$. Then, for each ${\bf k}$, the system will fully oscillate between these two states for arbitrarily small values of $\tilde\beta_{\bf k}$. 

Therefore, to obtain the correct physical values, we need to take the limit $\tilde\beta_{\bf k} \rightarrow 0$ of the general expression for $\bar q({\bf k})$ [Eq. (\ref{e-qbarrw})]. Then, we obtain
\begin{align}
    \lim_{\tilde\beta_{\bf k} \rightarrow 0} \bar q({\bf k}) = \frac{3}{8}.
    \label{e-qbarj0}
\end{align}

\paragraph{} $\mu_5=0$ implies $\alpha_{\bf k}=\tilde\beta_{\bf k}=0$, and then $\mu_3$ is also zero. Then the rotating wave Hamiltonian vanishes, and there is no dynamics: $q({\bf k}, t) = 1$. This occurs when $k_1=k_2=\pi$, and the number of modes satisfying this condition is $\sim N^{\frac{1}{3}}$.

To summarize: for generic driving conditions, \mbox{$5/16 \le \bar q({\bf k}) \le 3/8$}, for all values of ${\bf k}$ except those satisfying the condition $k_1+k_2=2\pi$ (the number of such ${\bf k}$-values $\sim N^{\frac{2}{3}}$); when the driving parameters are such that $J_0(J/\omega)=0$, then \mbox{$\bar q({\bf k}) = 3/8$}, for all values of ${\bf k}$. Since $Q$ is the average of $\bar q({\bf k})$ over ${\bf k}$, we conclude that, under all driving conditions, and as $N\rightarrow\infty$,
\begin{align}
    \frac{5}{16} \le Q \le \frac{3}{8}.
    \label{ea-Qbound2p}
\end{align}

\bibliography{reference}

\end{document}